# Magneto-structural phase transitions and two-dimensional spin waves in graphite

N Gheorghiu[1*], C R Ebbing[2], and T J Haugan[3]

[1]*UES, Inc., Dayton, OH 45432*

[2]*University of Dayton Research Institute, Dayton, Ohio 45469*

[3]*The Air Force Research Laboratory, Wright-Patterson AFB, Ohio 45433*

*Email: Nadina.Gheorghiu@yahoo.com

**Abstract.** We have previously found experimental evidence for several quantum phenomena in oxygen-ion implanted of hydrogenated graphite: ferromagnetism, antiferromagnetism, paramagentism, triplet superconductivity, Andreev states, Little-Parks oscillations, Lamb shift, Casimir effect, colossal magnetoresistance, and topologically-protected flat-energy bands [1-6]. Triplet superconductivity results in the formation of Josephson junctions, thus with potential of being used for spintronics applications in the critical area of quantum computing. In this paper, we are showing new experimental evidence for the formation of two-dimensional (2D) spin waves in oxygen-ion enriched and in hydrogenated highly oriented pyrolytic graphite. The temperature evolution of the remanent magnetization $M_{\text{rem}}(T)$ data confirms the formation of spin waves that follow the 2D Heisenberg model with a weak uniaxial anisotropy. In addition, the step-like features also found in the temperature dependence of the electrical resistivity between insulating and metallic states suggest several outstanding possibilities, such as a structural transition, triplet superconductivity, and chiral properties.

## 1. Introduction

As known, magnetic correlations are intrinsic to graphite. There are antiferromagnetic (AFM) correlations between unlike lattices (ABAB…) and weak ferromagnetic (FM) correlations between like sublattices (AAA… or BBB…). The presence of a critical number of vacancies/defects results in nucleation of ferrimagnetic islands that can cluster to form FM domains. A metamagnetic phase transition from an AFM martensite to a FM austenite phase and even evidence for SC was found in boron-carbon samples [7]. The existence of topological disorder in graphitic layers was theoretically predicted to lead to AFM, FM, and superconducting (SC) instabilities [8]. Moreover, sufficiently strong repulsive electro-electron (*e-e*) interactions favour the emergence of *p*-wave SC and can lead to HTSC. A review on the physics of graphite that includes phenomena discussed in this paper like magnetism, SC, and magnetic field induced phase transitions can be found in [9].

In this paper, following our previous work [4], we are presenting new experimental evidence for the formation of two-dimensional (2D) spin waves in oxygen-ion implanted and in hydrogenated highly oriented pyrolytic graphite (HOPG). The temperature evolution of the remanent magnetization $M_{\text{rem}}(T)$ data clearly shows the formation of spin waves that are generally described by the 2D Heisenberg model with a weak uniaxial anisotropy. In addition, we have found that beyond the region showing the 2D spin waves, $M_{\text{rem}}(T)$ goes through a magneto-structural transition from the AFM to the FM order that can be even discontinuous like in a first-order phase transition. In addition, step-like features are also found in the temperature dependence of the electrical resistivity, where changes from insulating to metallic states point to a structural transition, triplet superconductivity, and chiral properties.

## 2. Experimental Procedures

The carbon(C)-based materials used for this study are HOPG (particle size 10 μm) cylindrical-shaped samples (1 mm thick and 3 mm in diameter) and 2 mm × 2 mm × 1 mm graphite foil samples [10]. Within our extended project, some of the samples have been oxygen(O)-ion implanted at a two-dimensional (2D) concentration 2.24 x $10^{16}$ ions/cm$^2$. Details on the ion-implantation efforts can be found in [4]. The implantation energy was 70 keV and the implantation depth went up to 100% in graphite foils and up to 40% in the HOPG samples. Importantly, the ion implantation changes the hybridization of C atoms. Compared to perfect HOPG, some of the sp$^2$ bonds are lost and sp$^3$ bonds are formed instead, thus the ion implantation changes the sp$^2$/sp$^3$ ratio. In addition, mixed hybridized sp-sp$^2$ bonds are formed. The Raman spectra (figure 1) shows the presence of both sp$^3$ bonds revealed by the fullerenes' C60 peak and of mixed hybridized sp-sp$^2$ bonds revealed by the graphdiyne (GDY) peak. It is known that fullerenes, C onions, and other structures can be formed as the result of ion bombardment. This will become relevant and discussed in the next section. Other samples have been hydrogenated by intercalation of a hydrocarbon (an alkane, here octane $C_8H_{18}$). Graphite fibers have been either hydrogenated or O-implanted. More details can be found in [1,3]. The four-wire Van Der Pauw square-pattern technique was used for resistivity measurements. The quality of the silver electrical contacts was optically checked using an Olympus BX51 microscope. Magneto-transport and magnetization measurements were carried out in the 1.9 K − 300 K temperature (*T*) range and for magnetic fields of induction *B* up to 9 T using the Quantum Design Physical Properties Measurement System (PPMS) model 6500.

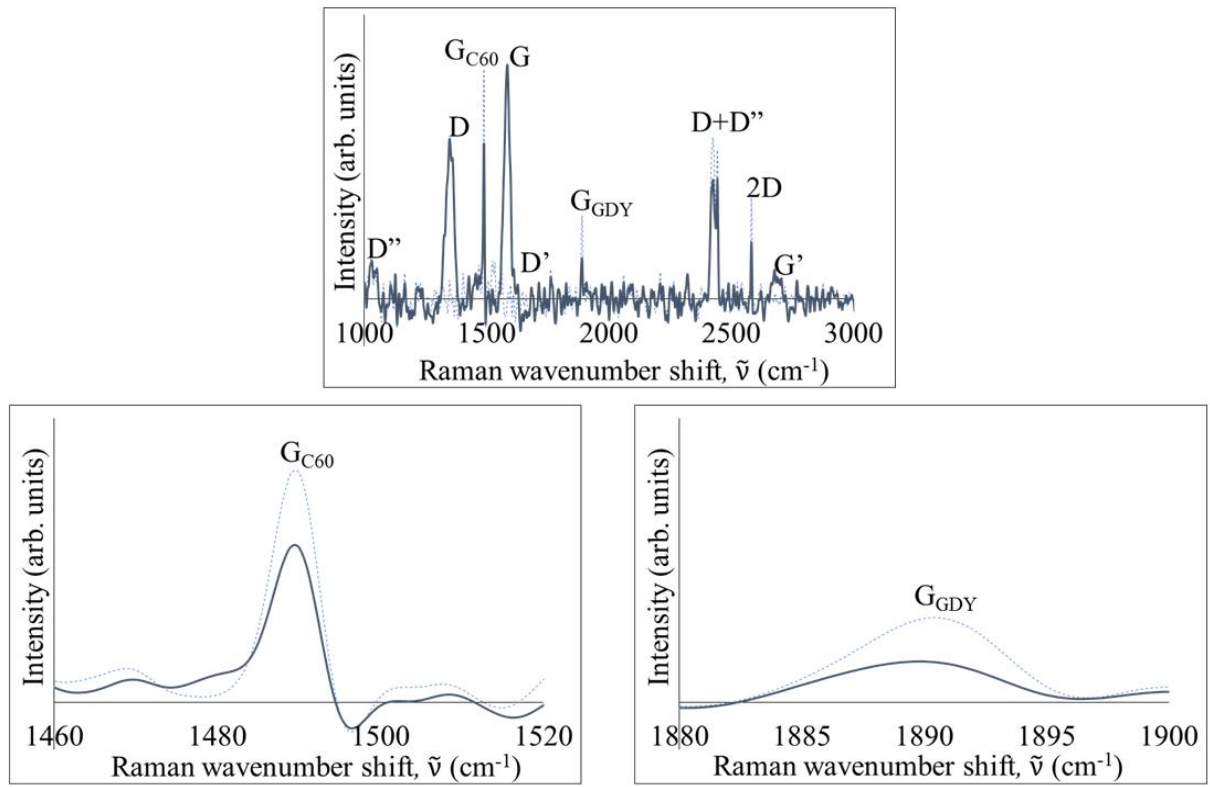

**Figure 1.** Raman spectra for HOPG: unmodified sample (solid line) vs. an O-implanted at a 2D concentration 2.24 x $10^{16}$ ions/cm$^2$ sample (dash line). The presence of fullerene C60 and graphdiyne (GDY) is discussed in the text.

## 3. Results and Discussion

Remanent magnetization data was taken after the sample was first magnetized at $T = 1.9$ K in high magnetic field of induction $B = \pm 9$ T. $T$-stabilized data was taken while the sample was warmed up to 300 K in zero field, $M_{remanent}(T, B = 0)$. Figure 2a shows the data for an HOPG sample. The remanent magnetization relaxes following a pattern not observed before in graphite. They are three distinctive features: a) a several-stage linear dependence of $M_{remanent}$ on $T$; b) a change of slope from $dM_{remanent}(T)/dT$ negative to positive, likely due to the nucleation and growth of ferrimagnetic islands; c) a step-like drop in $M_{remanent}$ at $T \cong 215$ K. The spin waves (SW) revealed by the $M_{remanent}(T, H = 0)$ experimental data are likely the result of reorientation of surface spins around the dislocations and/or domain pinning. The lowest energy in the spectrum for a spin wave is zero, thus a spin wave is easily formed on any lattice imperfection.

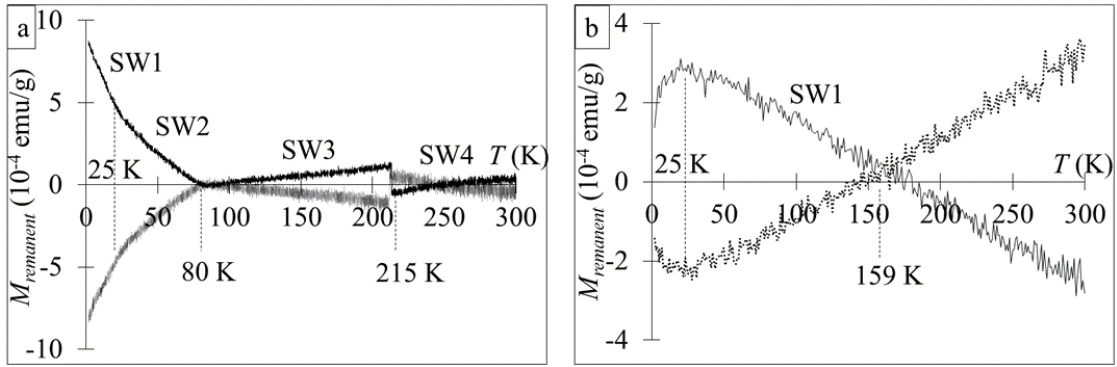

**Figure 2.** a) $T$-dependence of the remanent mass magnetization $M_{remanent}(T, H = 0)$ for a HOPG cylindrical sample after the application of a field of induction $B = 9$ T (solid line) and $B = -9$ T (round-dot line). b) The same measurement after the sample was annealed under a sodium lamp.

The spin waves can be described within the 2D Heisenberg model with a weak uniaxial anisotropy [11]. The normalized spin-wave magnetization in the anisotropic axis has the following low-$T$ dependence: $M_z^{sw} = 1 - T/T_c^{sw} - 2T^2/(T^*T_c^{sw}) - (2/3)(T/T_c^{sw})^3$, where $T^* = 4J$ ($J$ = the exchange coupling) and $T_c^{sw}$ is the critical $T$ for the spin-wave that is due to low-energy spin-wave excitations and it is given by $k_B T_c^{sw} = 2\pi J/K(D)$, where $K(\Delta)$ is the elliptic function with $\Delta$ the uniaxial anisotropy in the $z$ direction. Near the critical temperature $T_c$, the physics can be better described by a 2D Ising model, which should provide a good description of the spin-flip excitations. At low $T$, the normalized magnetization is given by: $M_z^{sw} \cong 1 - T/T_c^{sw}$.

Although the source for the magnetization data here is different than in [11], our data shows clearly that we are dealing with spin waves. Moreover, all spin waves observed here show practically a linear dependence. Figure 2a shows that $M_{remanent}(T)$ decreases with increasing $T$, it brakes (i.e., the spin wave changes slope) at $T_{SW1-SW2} \cong 25$ K, it is canceled at $T_{SW2-SW3} \cong 80$ K, and shows weak FM for higher $T$. A discontinuous (first-order) magneto-structural transition occurs at $T_{SW3-SW4} \cong 215$ K caused by the interaction between the magnetic and lattice orders, also reflecting the fact that some regions did not complete yet the transition from AFM to FM.

The observed step-like feature of the transition in $M_{rem}(T)$ might be also due to the coupling between the SC and the FM domains. The two critical values for $T$ are the found at the intersection of the almost linear $M_{remanent}(T)$ for the two low-$T$ spin waves: $T_c^{sw1} \cong 50$ K and $T_c^{sw2} \cong 80$ K, respectively. The former $T$ (50 K) was found before to be the SC transition temperature $T_c$ in O-implanted as well as hydrogenated graphitic, graphite-based or C-based samples, also the onset for insulating-SC metastability, Higgs-Anderson -like mode, in diamond-like C thin films, respectively, while the latter

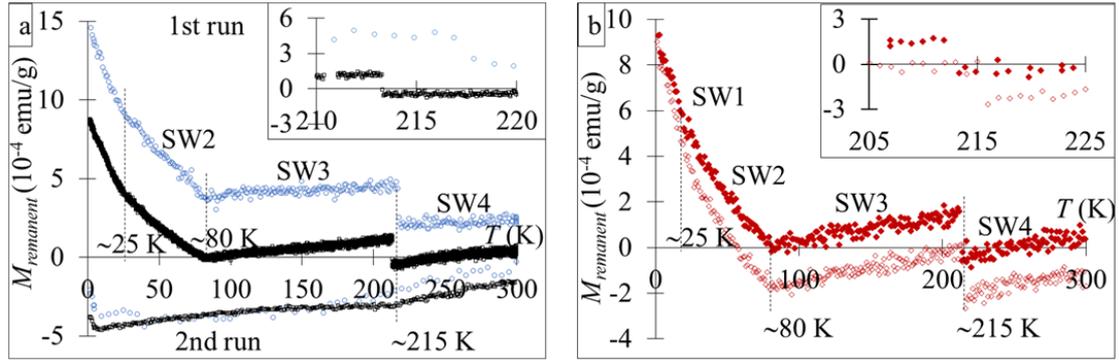

**Figure 3.** a) $M_{remanent}(T, H = 0)$ for HOPG cylindrical samples: as is (squares) and O-ion implanted at a dose 2.24 x $10^{16}$ ions/cm$^2$ (circles), respectively. b) $M_{remanent}(T, H = 0)$ for a hydrogenated cylindrical HOPG sample with no magnetic field applied (filled circles) and in small field $H$ = 25 Oe (circles), respectively.

(80 K) was the (possibly also Bogoliubov-deGennes) $T$ for the onset SC metastable state in amorphous-C thin films, respectively [1-6]. The spin waves found here are additional proof that magnetism and SC can coexist in these materials. Figure 2b shows the data obtained after the sample was annealed under a sodium lamp. As expected, the annealing erased some of the initial magnetic order, most significantly, the discontinuous step in the magnetization. A linear behavior is observed above $T \cong 25$ K corresponding to only one spin wave with $T_{SW1} \cong 159$ K.

The same features were observed for the modified HOPG samples. Figure 3a shows $M_{remanent}(T, H = 0)$ for a HOPG cylindrical sample that has been O-implanted at a dose 2.24 x $10^{16}$ ions/cm$^2$ vs. the unmodified HOPG sample. Again, four spin waves are observed. Notice that the PM is enhanced in the O-implanted sample. The second run shows no discontinuity in $M_{remanent}$ at $T \cong 215$ K due to the samples' annealing, at the same time, the DM reinstates (negative magnetization lines). We also see that the line for the O-implanted HOPG sample has an PM upturn at $T \cong 50$ K, which was found before as $T_c$ for entering the SC state in the O-implanted fiber [3], and thin films of O-implanted diamond-like carbon and O-implanted amorphous C [4].

Figure 3b shows $M_{remanent}(T, H = 0)$ for a hydrogenated cylindrical HOPG sample with no magnetic field applied (in red) and in small field $H$ = 25 Oe (in orange), respectively. The four spin waves are observed in both cases, with practically no effect from $H$. Here too, a discontinuous magneto-structural transition is observed at $T \cong 215$ K. Above $T \cong 50$ K, $M_{remanent}(T, H = 25$ Oe$)$ is negative. As known, graphite becomes more DM with increasing the strength of the applied magnetic field. We have found before that the hydrogenation of graphite leads to HTSC [1]. While 215 K is the melting point for octane, the step in $M_{remanent}(T \cong 215$ K, $H = 0)$ is observed also in the O-implanted sample, thus is clearly due to a magneto-structural transition in graphite. The presence of stacking faults can lead to 'earthquake-like' changes in graphite when under stress factors like $T$ and the magnetic field.

The discontinuous magneto-structural transition is not observed with the flexible graphite foil samples (figure 4a). As compared to the hard HOPG samples, the 'knee' at $T \cong 25$ K is now an inflection point. There are only two slightly different-slope spin waves in the flexible form of graphite, with the average slope for the two spin waves giving $T_{c,sw1\&sw2} \cong 58 - 60$ K. Remarkable, $M_{remanent}$ is not discontinuous at $T \cong 215$ K anymore. Instead, the sample shows a constant high-$T$ magnetization characteristic to its intrinsic FM. Thus, there is no magneto-structural transition in the flexible graphite foil. Figure 4b shows the $M_{remanent}(T, H = 0)$ for an unmodified graphite foil sample (black) vs. an O-implanted graphite foil sample (at a dose $2.24 \times 10^{16}$ ions/cm$^2$, blue) after the application of a field of induction $B = 9$ T. Notice that the O-ion implantation, which removes many of the sample's defects causing defect-induced magnetism, results in lower $M_{remanent}(T, H = 0)$ values.

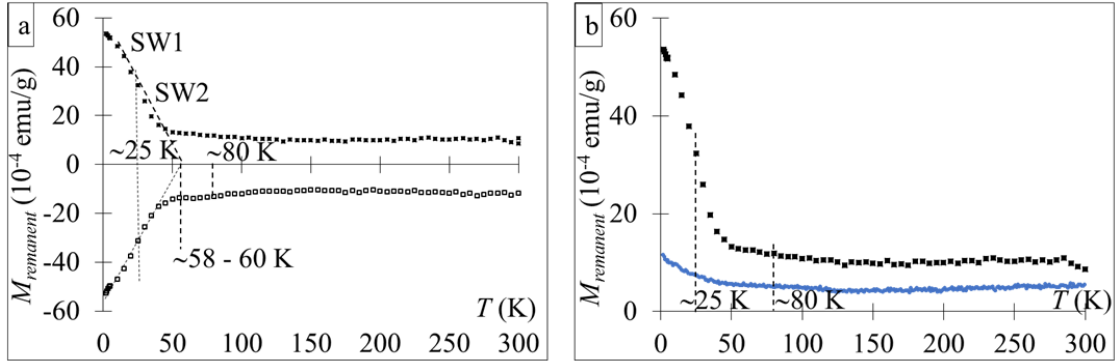

**Figure 4.** a) $M_{remanent}(T, H = 0)$ for an unmodified graphite foil sample after the application of a field of induction $B = 9$ T (filled squares) and $B = -9$ T (open squares), respectively. b) $M_{remanent}(T, H = 0)$ for an unmodified graphite foil sample (filled squares) vs. an O-implanted graphite foil sample at a dose $2.24 \times 10^{16}$ ions/cm$^2$ (filled circles) after the application of a field of induction $B = 9$ T.

Step-like features that might be caused by structural transitions were also observed in the (electric) transport data without magnetic field. The electrical resistivity (without magnetic field) for an O-implanted HOPG sample shows the presence of I-M transitions at $T \sim 280\text{-}290$ K and $T \sim 60$ K, respectively. There is also a metal-insulator (M-I) at $T \sim 40$ K (figure 5a). For bundles of graphitic fibers, I-M transitions are observed at $T \sim 260$ K, 210 K, 60 K, and 25 K, respectively. The metallic-like behavior observed with these bundles of C fibers, particularly the one below $T \simeq 260$ K, can be the result of parallel contributions of neighboring interfaces (graphene planes) to the electrical resistance.

In [4] we have suggested the fact that fullerene C60 clusters could be formed as a result of O-implantation. Indeed, the Raman spectrum confirmed the presence of C60 in the HOPG samples. This is in addition to HOPG being found to nurture on its surface self-assembling of highly-oriented chain structures of C60 molecules [15]. The heat treatment applied to obtain graphitic materials like fibers or the HOPG samples cut from extruded graphite rods leads to carbonization in the form of C60 fullerenes and their clusters. Interestingly, the C60 crystal clusters undergo a structural first-order phase transition from their room $T$ face-centered cubic structure to a simple cubic crystal structure at $T \sim 260$ K [12,13]. Thus, the M-I transition observed here at $T \sim 250$ K could be a structural phase transition. In hydrogenated graphite, the fullerenes can also encapsulate the H and possibly form two-fullerene cluster units [1]. Further, percolation between these hydrogenated fullerene clusters leads to SC.

I-M transitions were also observed for single fibers. Figure 6a shows the $T$-dependence of electrical resistivity, $\rho(T)$, for an unmodified T300 graphitic fiber. There are several step-like transitions: I-M at 250 K, 225 K, 175 K, and M-I transitions at 150 K, and 100 K, respectively. Figure 6b shows the $T$-dependent resistivity $\rho(T)$ for a hydrogenated T300 graphitic fiber. As the direct current $I$ increases, $\rho$ goes to lower values, i.e., the sample becomes more metallic. M-I transitions occur at $T \cong 215$ K and at $T \cong 230$ K. $T \cong 280$ K is remarkable here not only for the local minimum in $\rho$, also for the unconventional SC feature known as 'strange metal' observed above it, when strong electronic correlations result in $\rho(T) \propto T$. Due to the flexible nature of graphitic fibers, therefore not having stacking faults like in graphite, there should be another explanation for the step-like changes in $\rho(T)$. In this case, we have argued that structural/lattice modifications can also cause I-M transitions [2]. As known, a structural phase transition at a higher $T$ precedes the charge density wave order occurring at a lower $T$ in the SC state.

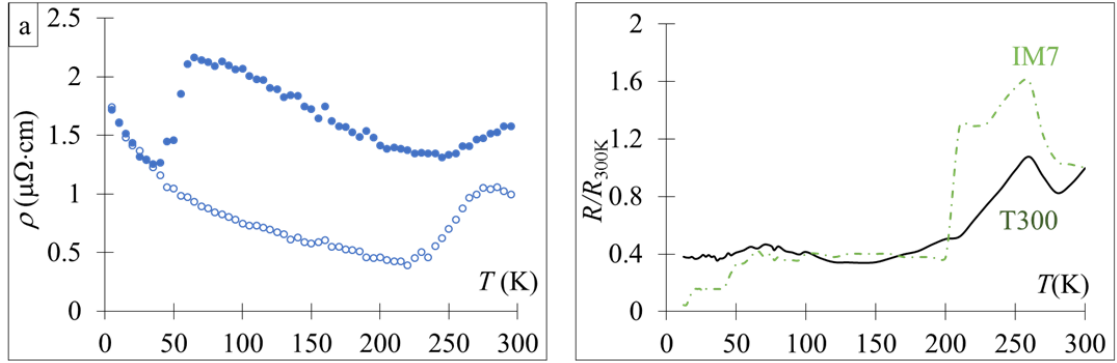

**Figure 5.** a) $T$-dependent resistivity $\rho(T)$ for a HOPG cylindrical sample that had been O-implanted at a dose 2.24 x $10^{16}$ ions/cm$^2$. As the $\rho$ values show, the sample becomes significantly more metallic for a direct current was $I = 5$ mA. Both cooling (filled circles) and warming (empty circles) cycles are shown. b) $T$-dependent resistance $R(T)$ data normalized to the value at $T = 300$ K for bundles of graphitic fibers T300 (1000, solid line) and IM7 fibers (12000, dash-dot line). The input direct current was $I = 20$ µA. Single fiber's diameter is 7.0 µm and 5.2 µm for the T300 fiber and the IM7 fiber, respectively.

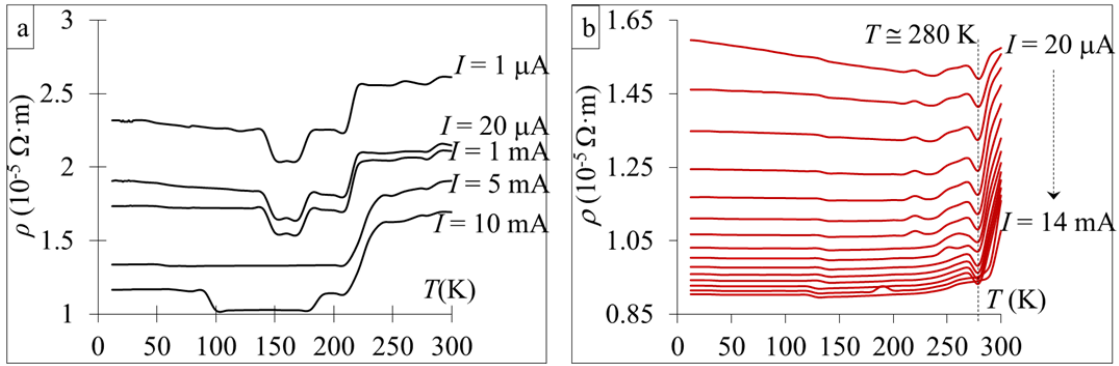

**Figure 6.** a) $T$-dependence of the electrical resistivity, $\rho(T)$, for an unmodified T300 graphitic fiber. b) $\rho(T)$ for a hydrogenated T300 graphitic fiber under different direct current inputs $I = 20$ µA and $I = 1$ mA to 14 mA in 1 mA steps, respectively.

The step-like structural changes observed for $T = 250$-$300$ K are remarkable for their possible relation to chirality [2]. Interestingly, while the C content is about 93%, the O-implanted polyacrylonitrile-based T300 graphitic fiber is made out of the same elements found in amino acids: H, N, C, and O. It has been suggested that biomolecular homochirality can be achieved through a D-amino acid→ L-amino acid as the $C_\alpha$-H bond would break and the H atom would become SC. Specifically, chirality among the twenty amino acids which make up the proteins may be a consequence of a phase transition which is analogous to that due to BCS SC [14].

Competing and coexisting charge, spin, and SC orders in graphite-based materials results in many interesting effects marked by their complex interplay. Thus, the magnetoresistance for a hydrogenated graphitic fiber reveals an unusual *T*-dependence (figure 7). While it appears that some SC phases survive to higher *T*, the main SC phase below $T_c \cong 50$ K is excitonic [1]. In parallel field, the magnetoresistance has extrema at $T \cong 140$ K, which is close to the BCS-type mean-field critical *T* for graphene, $T_c = 150$ K. We also mention that $T_c \cong 60$ K is the mean-field $T_c$ for SC correlations in the metallic-H multilayer graphene or in HOPG [16] and was also found as the spin glass critical *T* extracted from the Almeida-Thouless line [4]. The relevant *T* data points for $M_{remanent}(T, H = 0)$ data are seen here too: 25 K, 50 K, and 210-220 K, respectively.

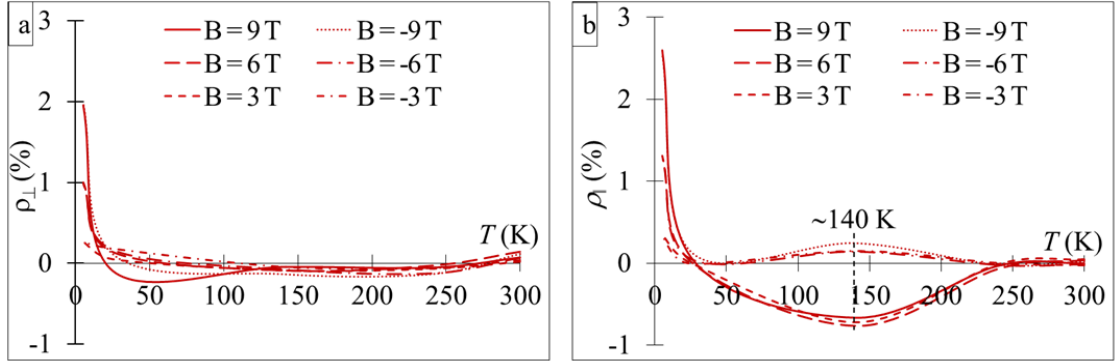

**Figure 7.** *T*-evolution of the magnetoresistance for a hydrogenated graphitic fiber. The fiber's orientation is: ⊥ = along the magnetic field, ∥ = along the fiber, respectively. *I* = 20 µA was applied.

It is possible that the SC is predominantly *s*-wave (singlet) pairing between about 50 K to 250 K and predominantly *p*-wave (triplet) pairing below 50 K and above 250 K. In [1], we have found for the hydrogenated graphite fiber at *T* = 130 K a small gap 11.6 µeV, about twice the Lamb shift 4.38 µeV value needed in the H atom for a split between the $2s_{1/2}$ and $2p_{1/2}$ energy levels. As known, the Lamb shift is related to the coexistence of SC and magnetism, including the coexistence with an incommensurate spin-density wave (magnon) at a structural transition to an anisotropic phase (in the *ab* plane).

SC is close to a magnetic instability. The spin-density wave and SC are features for strongly correlated electrons. AFM was observed below $T_{Néel} \cong 50$ K and FM survived up to room *T* [1]. Owning to the presence of spin carrying protons H+ introduced by the intercalation with an alkane, it is possible that the octane-intercalated/hydrogenated graphitic fiber is a chiral FMSC. Moreover, the parallel magnetic field can drive a topological phase transition with the formation of Majorana bound states, which are essential to the functioning of fault-tolerant quantum computing. Indeed, from nonlocal electrical differential conductance $G_{diff}$ measurements, we have found evidence of Andreev edge states and crossed Andreev conversion, moreover, the existence of topologically protected flat energy bands [1,6,17].

## 4. Conclusion

Charge, spin, and lattice orders are intertwined in quantum matter and can lead to HTSC. The structural, magnetic, conductive, and SC properties of 2D materials like graphite can be explored for the creation of integrated spintronic devices like FM-SC-FM spin valves which function based on the formation of the odd frequency triplet SC pair correlations. The possible existence of a Lamb shift in a hydrogenated graphitic fiber is also of particular importance. As known, the Lamb shift is central for the development of quantum electrodynamics. Small energy shifts like the Lamb shift or the Casimir are important for quantum computing. Understanding the many-faceted magnetism in graphite materials like the magneto-structural and lattice transitions revealed in this work, as well as their conductive and SC properties, are both practically necessary and fundamentally interesting research topics.

**Acknowledgments**
This work was supported by The Air Force Office of Scientific Research (AFOSR) for the LRIR #14RQ08COR & LRIR #18RQCOR100 and the Aerospace Systems Directorate (AFRL/RQ). Experimental data was acquired during the affiliation to UES, Inc. We acknowledge J.P. Murphy for the cryogenics and T.J. Bullard for discussions. N. Gheorghiu is especially grateful to G.Y. Panasyuk for his continuous support and inspiration.